\documentstyle[11pt,epsfig,amsmath,amssymb]{article}
\def\etal{{\it et al.}}

\begin{document}

\title{\bf Comment on `Search for new particles decaying into electron pairs of mass below 
100 MeV/$c^{2}$'}

\author{{F. W. N. de Boer\footnote{Corresponding author: e-mail:fokke@nikhef.nl} $\,$ and C.A.
Fields\footnote{Present address: 21 Rue des Lavandi\`eres, Caunes Minervois, 11160 France; e-mail:chris@hayfields-communications.com}}\\
{\it NIKHEF, Amsterdam, The Netherlands}}
\maketitle

\begin{abstract}
A re-analysis of data from electron-pair production following 160 A$\cdot$GeV $^{207}$Pb bombardment 
of nuclear emulsions indicates the production and decay of neutral particles of significantly 
lower invariant mass and shorter lifetimes than previously claimed (\textit{J. Phys. G: Nucl. Part. Phys.} \textbf{34} (2007) 129-138).
\end{abstract} 
 
\section{Introduction}

Jain and Singh [1] report a study of electron-pair production following 160 A$\cdot$GeV $^{207}$Pb bombardment of nuclear emulsions, in which the energy and opening angles of pairs produced at 
distances greater than 50 $\mu{}m$ from identified interaction vertices were analysed to determine 
the invariant masses of presumed neutral particles \textit{X} decaying by the $X \rightarrow e^{+}{}e^{-}$ channel.  Derived neutral particle masses from 2 to 84 MeV/c$^{2}$ were reported, with lifetimes ranging from $10^{-15}$ to $10^{-13}$ s (Fig. 1b, 2a, 3 and 4 of [1]).  Such large derived masses do not appear to be consistent with the total pair energy and opening angle data that are reported (Fig. 1f of [1]).  A re-analysis of the data presented in Fig. 1f indicates presumed neutral particle \textit{X} invariant masses of 1.5 to 21 MeV/c$^{2}$, with lifetimes between $10^{-16}$ and $10^{-14}$ s.  This mass and lifetime range are consistent with previous indications of light neutral particles decaying to $e^{+}{}e^{-}$ from cosmic ray [2-3], emulsion bombardment [4-9] and nuclear decay [10-16] data.

\section{Data and analysis}

\begin{figure}[hbt]
\centerline{\epsfig{file=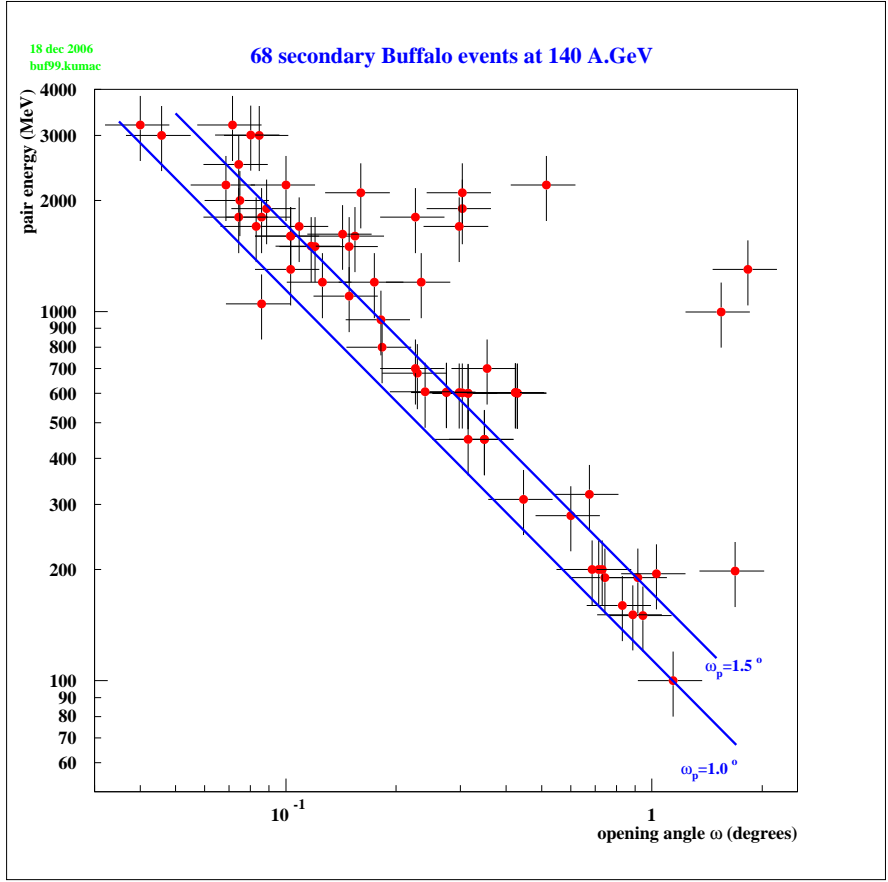,angle=0,width=4.6in}}
\caption{
Total pair energy $E_{tot}$ versus opening angle $\omega$ for the 62 events reported to be above the Borsellino line in Fig. 1f of [1].  Also shown are the Borsellino line $\omega_{P}$ = 4m$_{e}$c$^{2}$/E$_{tot}$ (in radians) and 1.5 times the Borsellino line.  Data were obtained manually from Fig. 1f of [1].  Uncertainties include the experimental uncertainties stated in [1] and the uncertainties associated with extracting the data from the published figures.
}
\label{fig:rawdata}
\end{figure}

Figure 1f of [1] shows a scatter plot of total electron-pair energy $E_{tot}$ versus opening angle $\theta$.  A total of 62 of the reported 1220 $e^{+}{}e^{-}$ pairs fall above Borsellino's [17] most probable opening angle $\omega_{P}$ at an invariant mass 1.02 MeV/c$^{2}$ as plotted together with the data in Fig. 1f of [1].  These 62 events are, therefore, the only events that can be interpreted as indicating the decays of massive particles into $e^{+}{}e^{-}$ pairs.  The total energies and opening angles of these 62 events are reproduced in Fig.~\ref{fig:rawdata}, together with plots of Borsellino's most probable opening angle $\omega_{P}$ and 1.5 times Borsellino's angle.

\begin{figure}[hbt]
\centerline{\epsfig{file=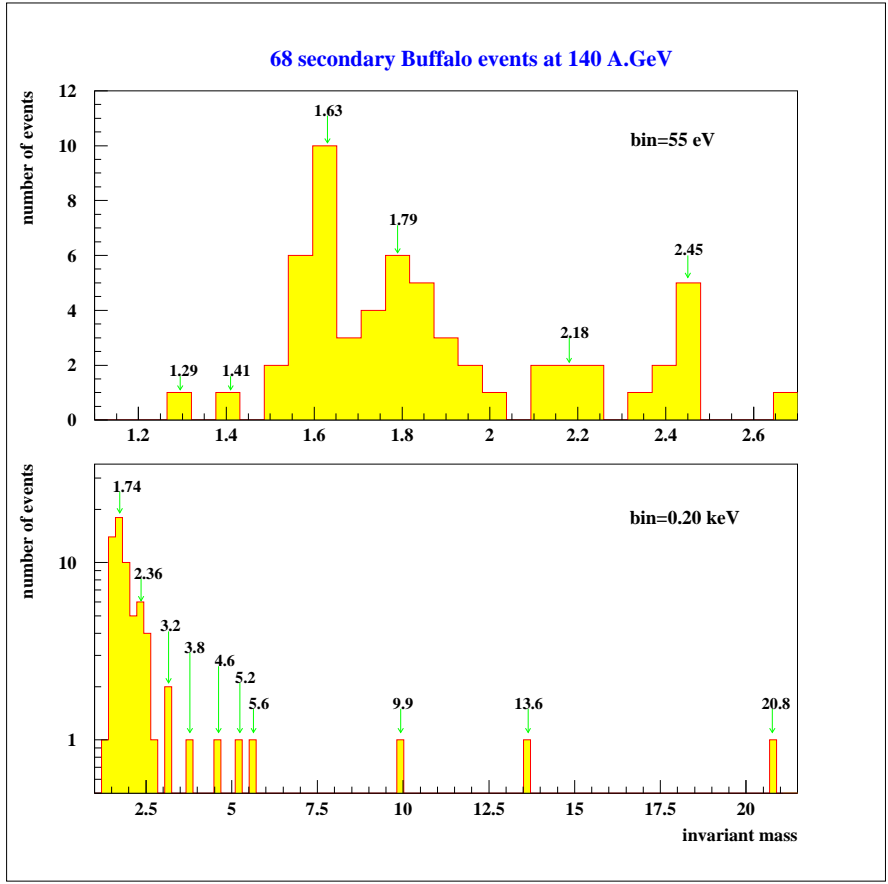,angle=0,width=4.6in}}
\caption{
Computed invariant masses $m_{X}$ of presumed neutral particles \textit{X} decaying to $e^{+}{}e^{-}$ pairs, as derived from Fig.~\ref{fig:rawdata} using Eqn. 1.  (a) 59 events in the range 1.5 MeV/c$^{2}$ $\leq m_{X} \leq$ 6.0 MeV/c$^{2}$.  (b) Full mass range, showing 3 events above 6 MeV/c$^{2}$.
}
\label{fig:massdata}
\end{figure}

The invariant mass $m_{X}$ for a presumed neutral particle \textit{X} decaying to $e^{+}{}e^{-}$ pairs was computed as:

\begin{equation}
m_{X}^{2} = 2m_{e}^{2} + 4E_{1}E_{2}sin^{2}(\omega{}/2)
\end{equation}

where $m_{e}$ is the electron rest mass and $E_{1}$ and $E_{2}$ are the pair electron energies, here taken to be equal on the basis of the very small energy divergences reported in Fig. 1e of [1].  The computed invariant masses are shown in Fig.~\ref{fig:massdata}.

The \textit{X}-particle lifetimes reported in [1] could not be confirmed directly, as the measured interaction vertex to pair vertex distances are not given and Figs 1b and 1f of [1] cannot be compared on a point-for point basis.  The highest density of events is reported [1] to be between 50 and 300 $\mu$m of an interaction vertex, and this distance range was used to calculate \textit{X}-particle lifetimes of $10^{-15}$ to $10^{-13}$ s [1].  From the mass range shown in Fig.~\ref{fig:massdata}, the \textit{X}-particle lifetime $\tau_{X}$ can be estimated to be between $1.7 \times 10^{-16}$ for a 1 MeV/c$^{2}$ \textit{X}-particle travelling 50 $\mu$m and $10^{-14}$ for a 10 MeV/c$^{2}$ \textit{X}-particle travelling 300 $\mu$m, i.e. the revised mass calculations yield a median lifetime of approximately $10^{-15}$ s, an order of magnitude lower than reported in [1]. 

\section{Discussion}

Evidence for neutral particles with masses less than 10 MeV decaying into $e^{+}{}e^{-}$ pairs has previously been reported from cosmic ray [2,3], emulsion bombardment [4-9] and nuclear decay [10-16] data.  The structure of these particles is unknown.  However, the angular distributions of $e^{+}{}e^{-}$ pairs emitted in a 10.96 MeV 0$^{-}$ $\rightarrow$ 0$^{+}$ magnetic monopole (M0) transition in $^{16}$O forbidden to both single-$\gamma$ and internal pair-conversion (IPC) modes indicates that an \textit{X}-boson of approximately 10 MeV/c$^{2}$ is the primary decay product in this transition [13-14].  Similarly, the analysis reported in [4,5] suggests that 1.1, 2.1 and 9 MeV/c$^{2}$ bosons are primary products of $^{12}$C and $^{22}$Ne bombardments of nuclear emulsions.  These candidate light neutral bosons have been tentatively interpreted as potential components of light dark matter [18].

The present re-analysis of the 160 A$\cdot$GeV $^{207}$Pb emulsion data reported by Jain and Singh [1] brings the majority of events observed in these reactions into a mass and lifetime range consistent with earlier reports of light neutral bosons decaying into $e^{+}{}e^{-}$ pairs.  A comprehensive re-analysis of data from multiple sources indicating the existence of such particles is currently under way [19].

\section{Acknowledgements}

We wish to acknowledge Keith Griffioen (William and Mary) and Andries van der Schaaf (University
of Zurich) for useful comments.

\section{Bibliography}

[1] Jain PL and Singh G 2007 \textit{J. Phys. G: Nucl. Part. Phys.} 34 129

\noindent
[2] Asikimori K \etal 1994 \textit{J. Phys. G: Nucl. Part. Phys.} 20 1257

\noindent
[3] Wilczynski H \etal 1997 \textit{Nucl Phys B} (Proc Suppl) 52 81

\noindent
[4] El-Nadi M and Badawi OE 1988 \textit{Phys. Rev. Lett.} 61 1271 

\noindent
[5] de Boer FWN and van Dantzig R 1988a \textit{Phys. Rev. Lett.} 61 1274; 1988b 
\textit{Phys. Rev. Lett.} 62 2639

\noindent
[6] El-Nadi M \etal 1966 \textit{Nuovo Cimento} A109 1517

\noindent
[7] Kamel S 1990 PhD Dissertation, Univ. Cairo

\noindent
[8] Kamel S 1996 \textit{Phys. Lett.} B368 291

\noindent
[9] El-Nagdy MS \etal 2007 AIP Conf. Proc. 888 249

\noindent
[10] de Boer FWN \etal 1996 \textit{Phys. Lett.} 388 235

\noindent
[11] de Boer FWN \etal 1997 \textit{J. Phys. G: Nucl. Part. Phys.} 27 L29

\noindent
[12] de Boer FWN \etal 1999 \textit{Nucl Phys B} (Proc. Suppl.) 72 189

\noindent
[13] de Boer FWN \etal 2001 \textit{J. Phys. G: Nucl. Part. Phys.} 27 L29 (arXiv:hep-ph/0101298) 

\noindent
[14] Stiebing KE \etal 2004 \textit{J. Phys. G: Nucl. Part. Phys.} 30 165 (arXiv:hep-ph/0311002)

\noindent
[15] Krasznahorkay A \etal 2006 \textit{Acta Phys. Polonica} 37 239 (arXiv:hep-ex/0510054)

\noindent
[16] Krasznahorkay A \etal 2008 \textit{Acta Phys. Polonica} 39 483

\noindent
[17] Borsellino A 1953 \textit{Phys. Rev.} 89 1023

\noindent
[18] Boehm C \etal 2004 Phys. Rev. Lett. 92 101301 (arXiv:astro-ph/0309686)

\noindent
[19] de Boer FWN \etal 2009 in prep.

\end{document}